\documentclass{emulateapj} \usepackage{apjfonts}
 
\slugcomment{ApJL, in press}    
    
\shorttitle{Diffuse light in clusters}    
\shortauthors{Murante, G. et al.}    
    
\begin{document}    
    
    
\title{The diffuse light in simulations of galaxy clusters}

\author{G. Murante\altaffilmark{1}, M. Arnaboldi\altaffilmark{1}, 
  O. Gerhard\altaffilmark{2}, S. Borgani\altaffilmark{3,4}, 
  L. M. Cheng\altaffilmark{3}, A. Diaferio\altaffilmark{5}, 
  K.~Dolag\altaffilmark{6}, L. Moscardini\altaffilmark{7}, 
  G. Tormen\altaffilmark{6}, 
  L. Tornatore\altaffilmark{3}, P. Tozzi\altaffilmark{8}} 
   
\altaffiltext{1}{INAF, Osservatorio Astronomico di Pino Torinese, 
  Strada Osservatorio 20, 10025 Pino Torinese, Italy 
  (murante@to.astro.it,arnaboldi@to.astro.it)} 
\altaffiltext{2}{Astronomisches Institut der Universitaet, CH-4102 
  Binningen, Switzerland (ortwin.gerhard@unibas.ch)} 
\altaffiltext{3}{Dipartimento di Astronomia dell'Universit\`a di Trieste, via   
  Tiepolo 11, I-34131 Trieste, Italy (borgani,cheng@ts.astro.it)}   
\altaffiltext{4} {   
  INFN -- National Institute for Nuclear Physics, Trieste, Italy}   
\altaffiltext{5}{ Dipartimento di Fisica Generale ``Amedeo Avogadro'',
  Universit\'a  degli Studi di Torino, Via Giuria 1, I-10125, Torino, Italy (diaferio@ph.unito.it) }   
\altaffiltext{6}{Dipartimento di Astronomia, Universit\`a di Padova, vicolo   
  dell'Osservatorio 2, I-35122 Padova, Italy (kdolag,tormen@pd.astro.it)}   
\altaffiltext{7}{Dipartimento di Astronomia, Universit\`a di Bologna, via Ranzani  1, I-40127 Bologna, Italy (moscardini@bo.astro.it)}   
\altaffiltext{8}{INAF, Osservatorio Astronomico di Trieste, via Tiepolo 11,   
  I-34131 Trieste, Italy (tozzi@ts.astro.it)}    
   
\begin{abstract}    
     We study the properties of the diffuse light in galaxy clusters
forming in a large hydrodynamical cosmological simulation of the
$\Lambda$CDM cosmology.  The simulation includes a model for radiative
cooling, star formation in dense cold gas, and feedback by SN-II
explosions.  We select clusters having mass $M>10^{14} h^{-1} M_\odot$
and study the spatial distribution of their star particles. While most
stellar light is concentrated in gravitationally bound galaxies
orbiting in the cluster potential, we find evidence for a substantial
diffuse component, which may account for the extended halos of light
observed around central cD galaxies.  We find that more massive
simulated clusters have a larger fraction of stars in the diffuse
light than the less massive ones.  The intracluster light is more
centrally concentrated than the galaxy light, and the stars in
the diffuse component are on average older than the stars in cluster
galaxies, supporting the view that the diffuse light is not a random
sampling of the stellar population in the cluster galaxies.  We thus
expect that at least $\sim 10\%$ of the stars in a cluster may be
distributed as intracluster light, largely hidden thus far due to its
very low surface brightness.
\end{abstract}    
    
    
\section{Introduction}\label{intro}    
The presence of diffuse `intracluster light' in galaxy groups and
clusters is now well established; observations by several groups
provide estimates of the fraction of diffuse light and its
distribution, using different techniques (see Arnaboldi 2003 for a
review). The fraction of stars contained in this space-filling
component seems to increase strongly with the density of the
environment: from loose groups ($ < 2\%$, Castro-Rodriguez et al.
2003; Durrell et al. 2003) to Virgo-like (~10\%; Feldmeier et
al. 2003; Arnaboldi et al. 2003) and rich clusters ($ \sim 20\%$ or
higher; Gonzalez et al. 2000, Feldmeier et al. 2002; Gal-Yam et
al. 2003). This correlation may represent an important clue for
understanding the mechanisms that produce intracluster (IC) light and drive
its evolution in the cluster environment.
    
Cosmological simulations of structure formation facilitate
studies of the diffuse light and its expected properties.
Dubinski (1998) constructed compound models
of disk galaxies and placed them into a partially evolved simulation
of cluster formation, allowing an evolutionary study of the dark
matter and stellar components independently.  Using an empirical
method to identify stellar tracer particles in high-resolution dark
matter (DM) simulations, Napolitano et al. (2003) studied a Virgo-like
cluster, finding evidence of a young dynamical age of the intracluster
component.  The main limitations in these approaches is the
restriction to collisionless dynamics.  
 
In this Letter, we analyze for the first time the IC light
formed in a cosmological hydrodynamical simulation including a
self-consistent model for star formation.  In this method, no
assumptions about the structural properties of the forming galaxies
need to be made, and the gradual formation process of the stars, as
well as their subsequent dynamical evolution in the non-linearly
evolving gravitational potential can be seen as a direct consequence
of the $\Lambda$CDM initial conditions. It is therefore of immediate
interest whether this theoretical formation scenario makes predictions
for IC light consistent with observations. Using a large
volume of $192^3\, h^{-3}{\rm Mpc}^3$, we can furthermore study a
statistically significant sample of clusters at $z=0$, and analyze the
correlations of properties of diffuse light with, e.g., cluster mass
and X-ray temperatures.
 
\section{Cosmological simulations}\label{Cs}    
We analyze the large scale cosmological hydrodynamical simulation
(LSCS) of a ``concordance'' $\Lambda$CDM model ($\Omega_m=0.3$,
$\Omega_\Lambda=0.7$, $\Omega_{\rm b}=0.019\,h^{-2}$, $h=0.7$,
$\sigma_8=0.8$) of Borgani et al.~(2004, B04). It was carried out
with the massively parallel Tree+SPH code {\small GADGET} (Springel et
al. 2001), using $480^3$ DM particles and as many gas particles.  For
the periodic cube of size $192\, h^{-1}$ Mpc, the mass resolution was
thus $m_{\rm dm}=4.62 \times 10^9\, h^{-1} M_\odot$ and $m_{\rm
gas}=6.93 \times 10^8 h^{-1} M_\odot$; the Plummer--equivalent
softening length was $\epsilon=7.5\, h^{-1}$ kpc (at $z=0$).  Besides
gravity and hydrodynamics, the simulation accounts for star formation
using a sub-resolution multi-phase model for the interstellar medium
(Springel \& Hernquist 2003), feedback from supernovae explosions
(including the effect of galactic outflows), radiative cooling of
the gas (assuming zero metallicity) and heating by a photo--ionizing,
uniform, time--dependent UV background (Haardt \& Madau 1999).
    
Clusters are identified at $z=0$ using a standard 
friends-of-friends algorithm, with a linking length of $0.15$ times 
the mean DM inter-particle separation. We identified 117 
clusters with $M>10^{14}h^{-1}M_\odot$. Cluster 
centers were placed at the position of the most bound particle 
belonging to each group. X--ray temperatures and masses 
were evaluated at the radius $R_{200}$, which encloses an 
average density of $200$ times the critical density.
A detailed study of the X--ray properties of our clusters, 
together with a full description of the numerical simulation can be
found in B04. 
    
This cosmological simulation shows an encouraging
agreement with some of the most important observed X-ray cluster
properties. However, there are also a number of discrepancies that
remain unaccounted for, as in other comparable numerical work.
For instance, the observed radial
temperature profiles in the cluster centers are not reproduced, and the
fraction of collapsed baryonic mass (``cold'' gas and stars) appears
still too large. While it is likely that the resolution of these
problems will require an improved treatment of the intracluster gas
physics, nonetheless this simulation represents a useful tool to study
the physical properties of the diffuse light.
    
\subsection{Presence of diffuse light}\label{skid}    
For all clusters identified in the cosmological simulation, we analyze
the stellar distribution in the cluster volume. To
compare with surface brightness measurements, we compute the
projected density of stars by integrating along a line-of-sight
(LOS).  We then extract the 2D--radial profile of each
cluster centered on its most bound particle, computing density
profiles in 100 shells for the star component from $0.05 R^{i}_{200}$
to $2 R^{i}_{200}$. A {\it stacked} profile is obtained by averaging
the shell densities with the same radius $R/R_{200}$ over all
clusters. For each cluster, the stars in galaxies other than the cD or
brightest cluster galaxies (BCGs) are not included.

For the galaxy identification we use the publicly available package
SKID (Stadel 2001 \footnote{
http://www-hpcc.astro.washington.edu/tools/skid.html}) to identify
self--bound gas and star particle groups within individual clusters.
We selected a scale of $20 h^{-1}$ kpc, comparable to our
physical force resolution, as the typical SKID length--scale
$\tau$. SKID groups together particles lying near local maxima of the
density field, as determined using DM, star and gas particles. Then
those star and gas particles which have a total energy $T+V>0$, where
$T$ is the kinetic energy and $V$ the {\it local} gravitational
energy, are removed from each group.  All particles in a sphere of
radius $2\tau$ are considered for evaluating the gravitational energy.
We discarded groups having less than 32 star particles.

\begin{figure}    
\plotone{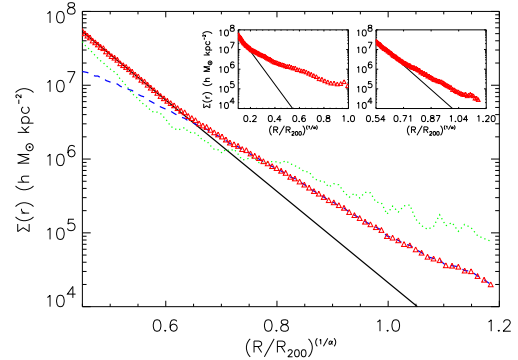} 
\caption{Schombert--like analysis on the {\it stacked} 2D radial
density profile (BCG + ICL) of clusters in the simulation
(triangles). The light excess is evident at large cluster radii. We
used an non--linear Levenberg--Marquardt fit to the function $ \log
\Sigma(r) = \log \Sigma_e -3.33 [(r/r_e)^{1/\alpha} -1]$ (solid line),
with best--fit parameters $\log \Sigma_e = 20.80$, $r_e =
0.005$, $\alpha = 3.66$. Also shown are the averaged 2D density profile
of stars in galaxies (dotted line) and in the field (dashed line), as
determined by SKID (see \P~\ref{skid}).  In the inserts, we show the
same analysis for the most luminous clusters with $T>4$ keV (left
panel), and for less luminous ones with $0<T<2$ keV (right panel).
The resulting best--fit parameters are respectively $\log \Sigma_e =
16.47$, $r_e = 0.11$, $\alpha = 1.24$ and $\log \Sigma_e = 23.11$,
$r_e = 0.00076$, $\alpha = 4.37$.  In the main plot and in the inserts
the unit $(R/R_{200})^{1/\alpha}$ refers to the $\alpha$ values given
by each Sersic profile.}
\label{fig1}    
\end{figure}

We then look for a halo stellar component in the stacked 2D profile,
following Schombert (1986), by checking whether the average radial
surface brightness profile curves upward in a $(\mu,R^{1/\alpha})$
diagram, over significant intervals in $R^{1/\alpha}$.  We fit a
Sersic law to the inner parts of the {\it stacked} surface brightness
profile, in a range of radii from the center out to a radius where the
surface density is about one third of the central value.  The result
is shown in Fig.~\ref{fig1}: a light excess to the inner Sersic
profile is evident for radii $R/R_{200} > 0.18 $.  The deviation at
large $R$ of the {\it stacked} surface brightness profile from the
Sersic's law is interpreted as being due to an extensive luminous
halo. In observed cDs this occurs at $R \sim 50 \div 80$ kpc, which is
smaller than our measured value. Previous studies of cluster evolution
(Dubinski 1998, Napolitano et al. 2003) produced clusters with BCGs
whose density profile followed a de Vaucoleurs' law at all radii. In
our simulations, where gas, stars and DM particles are followed
self-consistently during structure formation, cD halos do form.

According to Dressler (1979) and Kormendy (1980) the 
upturn in the cD surface brightness
$(\mu,R^{1/\alpha})$ plot, indicating the additional luminous halo,
would occur at the projected radius where the stars become unbound
from the central elliptical galaxy and orbit in the cluster
potential. The velocity dispersion profile would then rise at the
radius where the change of slope takes place. A similar effect is
observed in NGC~1399, a nearby cD galaxy (Arnaboldi et al. 1994,
Napolitano et al. 2002).
    
In a cosmological simulation, we have the phase space information for
all particles; thus, we can study the dynamical behavior of those
particles which populate the outer halo of cD galaxies.  We will refer
to all star particles grouped to any substructures by SKID as
``bound'', while the others will be named ``unbound''. The distinction
between stars bound to a galaxy and unbound stars is workable except
for the cluster center, where the most bound IC stars and the stars of
the central cD fall in the same part of phase-space.  

\begin{figure}    
\plotone{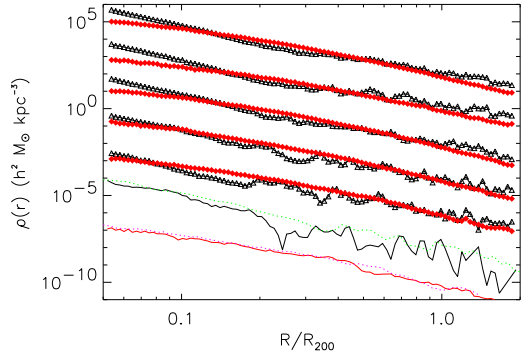}    
\caption{ 3D radial density profiles of the bound (open triangles) and
unbound (filled diamonds) stars, for clusters divided in X--ray
temperature classes. From top to bottom, the lines are for: global
average, $T\!<\!2$ keV, $2\!<\!T\!<\!3$ keV, $3\!<\!T\!<\!4$ keV,
$T\!>\!4$ keV. The last four couples of density profiles have been divided
by $10^2, 10^4, 10^6, 10^8$ for clarity. The radius where the bound
component dominates shifts toward smaller fractions of the average
$R_{200}$, while the temperature, and thus the mass of clusters,
increases.  We also show the 3D radial density profiles of one cluster
compared with its re-simulation R10 (see text). Solid lines refer to
R10, dashed lines to the LSCS cluster. All 
density profiles for ``unbound'' and ``bound'' stars have been divided by
$7\cdot 10^9$ and $10^{12}$, respectively.  }
\label{fig2}    
\end{figure}

Once the stellar particles in a given cluster are flagged, we build up
the global 3D radial profiles for the ``bound'' and the ``unbound''
stars.  Then we group all the clusters in classes with different X-ray
temperature, and derive the average 3D stellar density profiles for
the ``bound'' and ``unbound'' components. Fig.~\ref{fig2} shows that
the ``unbound'' stars have a shallower radial profile than the
``bound'' component, in the range of radii from $0.05 R_{200}$ to
$0.3R_{200}$. These stars are responsible for the additional light
detected in cD halos, and build up the diffuse light in our clusters.

\section{Properties of diffuse light}    
The size of our simulated volume allows us to study the physical
properties of the IC stellar light in clusters statistically, and
explore their dependence on the total mass and/or the X-ray
temperatures of clusters. For all the
clusters selected in our database, we evaluate the logarithmic slope
of the 3D profiles in a range of radii around $R_*$, where $R_*$ is
the cluster radius where $\rho_{*,\rm bound} = \rho_{*,\rm
unbound}$. In the radial interval $R_*/1.5 \!<\! R \!<\! 1.5\cdot R_*$, the
profile slopes for the diffuse stars are in the range (-1, -3), while
the profile slopes for the ``bound'' stars are in the range
(-3,-5). Here the diffuse component has a shallower slope than the
``bound'' stars belonging to the central cD. At larger radii, the
``bound'' stars are those in cluster galaxies, and they become
dominant over the more centrally concentrated IC light, see
Fig.~\ref{fig1} and Fig.~\ref{fig2}.
 
We also investigated whether any trends are present with cluster mass,
either for the slope or for $R_*$, but no significant trend was
detected.
    
\begin{figure}    
\plotone{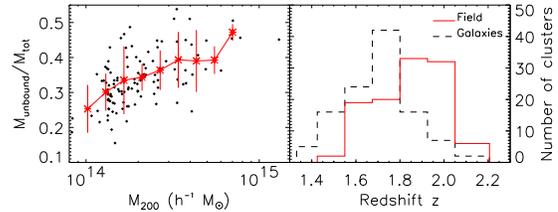}   
\caption{Left: Fraction of stellar mass in diffuse light vs.\ cluster
mass. Dots are for clusters in the simulated volume; asterisks show
the average values of this fraction in 9 mass bins with
errorbars. Right panel: histograms of clusters over mean formation
redshift, of their respective bound (dashed) and IC star particles
(solid line). Mean formation redshifts are evaluated for each cluster as
the average on the formation redshift of each star particle. }
\label{fig3}    
\end{figure}

Another relevant quantity is the fraction $f$ of diffuse vs. total
mass in stars: this quantity is a function of the cluster mass, with
the more massive clusters having a larger fraction of IC light,
see Fig.~\ref{fig3}.  We found that the amplitude of the correlation
between the fraction of diffuse vs. total stellar mass depends on the
value of the SKID length parameter $\tau$, but the slope of this
correlation is almost unchanged when $\tau$ changes.  This is
connected with the difficulty of separating the stars in the cD from
the IC stars near the cluster center. Large values of $\tau$ increase the
number of stars in galaxies, but the fraction of unbound stars never
drops to zero. In the most extreme case we tested ($\tau=40 h^{-1}$
kpc), only few of our 117 clusters drop to $f \!<\! 10\%$. These objects
are in the low--mass range ($M\!<\!2\cdot10^{14}h^{-1}M_\odot$).

We checked that $\tau=20 h^{-1}$kpc ensures a clear dynamical
separation between ``bound'' and ``unbound'' stars. The ``unbound''
stars have a 3D velocity dispersion comparable to that of the DM
particles, while ``bound'' particles have a smaller velocity
dispersion. Typical values for the most massive cluster in our LSCS
are $\sigma_{\rm unbound} \simeq \sigma_{\rm DM} \approx 3000$
kms$^{-1}$, $\sigma_{\rm bound} \approx 800$ kms$^{-1}$ at $R=100
h^{-1}$ kpc.

The simulation also records the age of formation of each star
particle.  In Fig. ~\ref{fig3} we investigate the age distribution of
the IC stellar components: stars in the diffuse component
formed from their parent gas particles at an earlier average redshift
($z \approx 1.9$), than ``bound'' stars ($z \approx 1.7$), with no
evidence for a dependence on cluster mass.  Our cosmological
simulation predicts that the stars in the IC component are older.

Two numerical effects can influence our results: the numerical
resolution, and the parameter $\tau$ for substructure identification.
To study the first effect, we simulated a cluster with total mass $2.9
\cdot 10^{14} h^{-1}M_\odot$, and a mass resolution increased by a
factor 3 and 10 respectively (R3 and R10 runs, here below). Softening
was rescaled as $m^{-1/3}$.  Initial conditions for these simulations
were generated using the ``zoomed initial conditions'' technique
(Tormen et al. 1997), which increases the resolution in the cluster
Lagrangian region while maintaining a coarse sampling of the
surrounding structures to account for their tidal field.  We find that
the amplitude and slopes of the density profiles for both bound and
unbound stars are almost unchanged at radii as large as $\approx 200
h^{-1}$ kpc (see Fig.~\ref{fig2}). The lower resolution in our
simulation may lead to an enhancement of the unbound population
because of numerical over-merging. However, when we increase the
resolution by a factor 10 in mass and $\approx 3$ in force, the
fraction $f$ does not change substantially: we find $f=0.41$ in LSCS,
$f=0.43$ for R3, and $f=0.38$ for R10 for the same cluster, suggesting
that the numerical resolution has only a small effect. Moreover, a
recent analysis of simulations of clusters having similar resolution
to our R3 (Sommer--Larsen et al.~2004), has independently confirmed 
the $f$ value reported here.
 
The SKID length parameter does influence the behaviour of both
``bound'' and ``unbound'' components in a number of expected trends,
i.e., the mass of the cD galaxy and the value of $R_*$ increase.
However, even when $\tau=40 h^{-1}$ kpc, the fraction of IC light
is not smaller than $0.1$ in the less massive clusters. Among those
$\tau$ values that we checked, $\tau=20 h^{-1}$ kpc ensures the best
dynamical separation between the two stellar components at the cluster
centers.

\section{Discussion}    

We used a cosmological simulation of $(192 h^{-1}{\rm Mpc})^3$ to
study the statistical properties of the IC light in clusters of
galaxies and the dependence of its physical properties on cluster mass
and X-ray temperature. These predictions can be tested against known
properties of cD halos and used to plan observational tests to
understand the physical properties of IC light.
 
The presence of the IC component is evident when the
whole distribution of stars in the simulated clusters is analysed in a
way similar to Schombert's (1986) photometry of BCGs. 
Galaxies at the center of our simulated clusters have
surface-brightness profiles which turn strongly upward in a
$(\mu,R^{1/\alpha})$ plot. This light excess can be explained as
IC stars orbiting in the cluster potential. 
Integrating its density distribution along the LOS,
the slopes from our simulations are in agreement with those observed
for the surface brightness profiles of the diffuse light in nearby
clusters. In the Coma cluster, Bernstein et al. (1995) parametrize the
surface brightness as $r^\beta$ and find that the diffuse light is
best fit by $\beta=-1.3\pm0.1$. In the Fornax cluster, the surface
brightness profile of the cD envelope of NGC 1399 follows a power law
of the form $\propto r^\beta$ with $\beta=-1.5$ (Bicknell et
al. 1989).

At large cluster radii, the surface brightness profile of the IC
light appears more centrally concentrated than the surface brightness
profile of cluster galaxies (see Fig.~\ref{fig1} and Fig.~\ref{fig2}).
From the simulations we also obtained the redshifts $z_{form}$ at
which the stars formed: those in the IC component have a
$z_{form}$ distribution which differs from that in cluster galaxies,
see Fig.~\ref{fig3}. The ``unbound'' stars are formed earlier than the
stars in galaxies. The prediction for an old stars' age in the diffuse
component agrees with the HST observation of the IRGB stars in the
Virgo IC field, e.g. $t> 2$Gyr (Durrell et al. 2002), and points
toward the early tidal interactions as the preferred formation process
for the IC light.  The different age and spatial distribution of the stars
in the diffuse component indicate that it is a stellar population that
is not a random sampling of the stellar populations in cluster
galaxies.

The more massive clusters have the largest fraction $f$ of diffuse light
(Fig.~\ref{fig3}). It is $f>0.1$ for cluster masses $M >
10^{14}h^{-1}M_\odot$.  Our simulations may thus explain the low
inferred star-formation efficiency in clusters vs. less massive
structures (David 1997). If only the {\it bound} stellar mass is
accounted for in the ratio of the total cluster stellar mass vs.\
cluster gas mass in our LSCS, then this ratio decreases from groups to
rich clusters. The observational trend would then be reproduced in the
simulation.  Similarly, the disagreement
found between the amount of stars produced in clusters in our LSCS and
in observed clusters (see B04) is less severe, if an IC component
is present in real clusters and has been systematically neglected when
evaluating their internal stellar mass budget.

The main result of this work is that large cosmological hydrodynamical
simulations are in qualitative agreement with the observed properties
of diffuse light in galaxy clusters. A quantitative assessment will
require additional numerical efforts and more observations.  A
detailed study of the dynamical history of the unbound stellar
population in our simulation will be presented in a forthcoming work.

\acknowledgments Simulations were performed on the CINECA IBM-SP4 
supercomputing facility, INAF grant for numerical Key--Project ``A 
Tree+SPH High--Resolution Simulation of the Cosmic Web''. This work is 
also supported by the INAF grant for national project (P.I. MA) and by 
the PD51 INFN grant.  OG 
thanks the Swiss Nationalfonds for support. KD acknowledges support by 
a Marie Curie fellowship of the European Community program ``Human 
Potential'' under contract number MCFI-2001-01221.  We thank 
V. Springel for carefully reading the manuscript prior to submission, 
and acknowledge A. Burkert, S. Bonometto, E. D'Onghia, A. Klypin and
J. Sommer-Larsen for useful discussions.


\begin{thebibliography}{}    
\bibitem[Arnaboldi et al. (2003)]{arn03+} Arnaboldi, M. et al. 2003, 
  \aj, 125, 514 
\bibitem[Arnaboldi (2003)]{arn03} Arnaboldi, M. 2003, ASP Conf. Series,    
  217 in press (astro-ph/0310143)   
\bibitem[Arnaboldi et al. (1994)]{arn+94} Arnaboldi, M., Freeman,   
 K. C., Hui, X., Capaccioli, M., \& Ford, H. 1994, ESO Messenger, 76, 40   
\bibitem[Bernstein et al. 1995]{brn95+} Bernstein, G. M., Nichol, 
 R. C., Tyson, J. A., Ulmer, M. P., \& Wittman, D.  1995, \aj, 110, 1507 
\bibitem[Bicknell et al. 1989]{bick89+}Bicknell, G. V., Bruce, 
 T. E. G., Carter, D., \& Killeen, N. E. B.  1989, \apj, 336, 639 
\bibitem[Borgani]{553} Borgani, S., Murante, G., Springel, V., Diaferio,   
A., Dolag, K., Moscardini, L., Tormen, G., Tornatore, l., \& Tozzi,   
P., 2004, \mnras, 348, 1078   
\bibitem[Castro-Rodriguez et al. (2003)]{castro03+}Castro-Rodriguez, 
  N., et al. 2003, \aap, 405, 803 
\bibitem[David (1997)]{david97} David, L.P. 1997, \apj, 484, L11   
\bibitem[Dressler (1979)]{dres79} Dressler, A. 1979, \apj, 236, 351   
\bibitem[Dolag et al. (2004)]{dolag04} Dolag, K., Jubelgas, M., Springel, V., 
  Borgani, S., \& Rasia, E., 2004, ApJ, submitted 
\bibitem[Dubinsky (1998)]{dub98} Dubinski, J. 1998, \apj, 502, 141    
\bibitem[Durrell et al. (2002)]{durr02_a} Durrell, P., et al. 2002, 
  \apj,  570, 119 
\bibitem[Durrell et al. (2003)]{durr02_b} Durrell, P., et al. 2003, ASP
  Conf. Series, 217 in press (astro-ph/0311130)
\bibitem[Feldmeier et al. 2002]{feld02+}Feldmeier, J.J., et al. 2002, 
  \apj, 575, 779 
\bibitem[Feldmeier et al. 2003]{feld03+}Feldmeier, J.J., et al. 2003, 
  \apjs, 145, 65 
\bibitem[Ferguson et al. (1998)]{ferg98} Ferguson, H., Tanvir, N. R.,    
\& von Hippel, T. 1998, Nature, 391, 461  
\bibitem[Gal-Yam et al. 2003]{gam03+}Gal-Yam, A., et al. 2003, \aj, 
125, 1087   
\bibitem[Gonzalez et al. (2000)]{gonz+0}Gonzalez, A.H., et al. 2000, 
  \apj, 536, 561 
\bibitem[Kormendy (1980)]{kor80} Kormendy, J. 1980. ESO workshop on   
  Two dimentional Photometry, ed. P. Crane \& K. Kjar   
  (Leiden:Sterrenacht Leiden) 191   
\bibitem[Haardt \& Madau (1999)]{hm99} Haart, F., \& Madau, P. 1999, \apj,   
  461, 20   
\bibitem[Napolitano et al. (2003)]{nap03} Napolitano, N.R., et    
  al. 2003,  \apj, 594, 172   
\bibitem[Napolitano et al. (2002)]{nap02} Napolitano, N.R., Arnaboldi, \&   
  M., Capaccioli, M. 2002, \aap, 383, 791    
\bibitem[Schombert (1986)]{scho86} Schombert, J. 1986, \apjs, 60, 603    
\bibitem[]{}Sommer--Larsen, J., Romeo, A.D., \& Portinari, L., \mnras,
submitted, astro-ph/0403282
\bibitem[]{587}Springel V., \& Hernquist L., 2003, \mnras, 339, 289    
\bibitem[]{588}Springel, V., Yoshida, N., \& White, S.D.M., 2001, NewA,   
  6, 79    
\bibitem[]{590}Stadel, J., 2001, Ph.D. Thesis, Washington   
\bibitem[]{591}Tormen, G., Bouchet, F. R., \& White, S.D.M., 1997,
\mnras, 286, 865
\bibitem[]{593}Tormen, G., Moscardini, L., \& Yoshida, N., 2004, \mnras,
in press (astro-ph/0304375)
\end{thebibliography}
\end{document}